\documentclass[10pt,aps,prl,twocolumn,superscriptaddress,numerical]{revtex4-2}
\usepackage[utf8]{inputenc}
\usepackage{amsmath,amssymb,amsthm}
\usepackage[english]{babel}
\usepackage[T1]{fontenc}
\usepackage{MnSymbol}
\usepackage{amsmath,dsfont,amsfonts,bbm,pifont}
\usepackage[colorlinks,
citecolor=purple,
linkcolor=purple,
urlcolor=purple]{hyperref}
\usepackage{tikz}
\usepackage{lipsum}
\usepackage{dsfont}
\usepackage{physics}
\usepackage{braket}
\usepackage{orcidlink}
\usepackage{multirow}
\usepackage[export]{adjustbox}
\usepackage[dvipsnames,svgnames]{xcolor}
\usepackage{color}
\usepackage{booktabs}
\usepackage{lipsum}
\graphicspath{{images}}

\newcommand{\Id}{\mathbbm{1}}

\newcommand{\PauliZ}{{\text{Z}}}

\newtheorem*{theorem}{Theorem}
\newtheorem*{prop}{Proposition}

\DeclareMathOperator{\Betad}{Beta}

\newcommand{\E}{\mathbb{E}}
\newcommand{\Var}{\text{Var}}
\newcommand{\Nh}{\bar N}

\newcommand{\eqd}{\overset{d}{=}}

\usetikzlibrary{arrows.meta}

\makeatletter
\renewcommand\paragraph{\@startsection{paragraph}{4}{\parindent}%
  {3.25ex \@plus1ex \@minus.2ex}%
  {-1em}%
  {\normalfont\normalsize\bfseries\itshape}}
\makeatother

\begin{document}

\title{Fermionic Porter–Thomas statistics and their emergence in random matchgate circuits}

\author{Guglielmo Lami~\orcidlink{0000-0002-1778-7263}}
\affiliation{Laboratoire de Physique Th\'eorique et Mod\'elisation, CNRS UMR 8089, CY Cergy Paris Universit\'e, 95302 Cergy-Pontoise Cedex, France}
\affiliation{JEIP, UAR 3573 CNRS, Collège de France, PSL Research University,
11 Place Marcelin Berthelot, 75321 Paris Cedex 05, France}

\begin{abstract}
For Haar-random quantum states, measurement probabilities follow the universal Porter--Thomas distribution. This implies anticoncentration, i.e.\ the spreading of probability over exponentially many outcomes, a key property underlying random-circuit-sampling experiments. We derive the exact analogue of the Porter--Thomas law for random fermionic Gaussian states, both with and without particle-number conservation. We also consider Clifford Gaussian states, for which the problem maps onto a simple model of arcs and loops. For large systems the probabilities become log-normally distributed, with moments growing polynomially in system size. Free-fermion states thus anticoncentrate only up to polynomial factors. We then show how this fermionic Porter--Thomas law emerges in random one-dimensional brickwork matchgate circuits. A simple single-arc approximation accurately describes the full distribution of probabilities, revealing an early diffusive regime followed by a slow relaxation on times scaling as the square of the system size. We rigorously prove that depths growing almost quadratically with system size are required for the collision probability to converge. Finally, we show that ideal free-fermion sampling yields large linear cross-entropy scores, and propose a log-based benchmark with an exact fermionic normalization.
\end{abstract}

\maketitle
\section{Introduction}

\emph{Fermionic Gaussian states} (FGS), the states of free fermionic particles, are fully characterized by their two-point correlation functions, and can therefore be simulated efficiently on a classical computer~\cite{Surace_2022}. This has made them a standard tool in condensed matter physics and quantum chemistry~\cite{Fradkin_2013,Giuliani_Vignale_2005,Szabo_1996,Mcardle_2020}, 
and a testing ground for many-body dynamics in clean~\cite{Rigol_2007,Essler_2016,Calabrese_2005,Alba_2017}, disordered~\cite{PhysRev.109.1492,Evers_2008} and stochastic~\cite{barraquand2026introductionquantumexclusionprocesses,PhysRevLett.123.080601} settings.
In quantum computation, free-fermion circuits, known as matchgate circuits~\cite{Valiant_2001,Terhal_2002,Jozsa_2008} or fermionic linear optics~\cite{Bravyi_2004}, are the paradigmatic example of classically simulable quantum dynamics and a controlled setting to probe what separates easy from hard computation~\cite{Bittel_2025,Mele_2025,Oszmaniec_2022,Cao_2026}.
Here, we focus on a fundamental property of FGS: the distribution of their outcome probabilities in the occupation-number basis. Its moments are the inverse participation ratios, a standard measure of how a state spreads over a basis, widely used to diagnose localization~\cite{Evers_2008,Murphy_2011,De_Luca_2013}. 

For Haar-random states on the full Hilbert space, this distribution takes the universal \emph{Porter--Thomas} (PT) form~\cite{Porter_Thomas_1956,Boixo_2018}. In particular, they \emph{anticoncentrate}: the probability is spread over many outcomes, rather than concentrated on a few. PT statistics and anticoncentration underlie quantum-advantage experiments based on the hardness of random circuit sampling~\cite{Boixo_2018,Arute_2019,sauliere2026observationuniversalquantumchaos}. They are reached at shallow depth by random quantum circuits~\cite{Dalzell_2022,Turkeshi_2024_1}, and the way they are approached is itself universal~\cite{xl16-cdy9,lkwg-4dbt,kl64-xnsz,PhysRevLett.134.010401}.
Anticoncentration has also been studied for restricted gate sets, particularly Clifford gates~\cite{w6l4-mh6k,p8dn-glcw}. For free-fermion systems the situation is different. Matchgate circuits are classically simulable, and sampling hardness arises only when they act on non-Gaussian ``magic'' input states~\cite{Hebenstreit_2019,Oszmaniec_2022,Mathur_2026,Cao_2026}, a case in which anticoncentration has been proven~\cite{Oszmaniec_2022,Mathur_2026,Cao_2026}. For Gaussian inputs, instead, the output statistics have not yet been characterized, nor has it been established whether, and to what extent, they anticoncentrate, a question left open in Ref.~\cite{Oszmaniec_2022}. This gap has practical consequences: the cross-entropy benchmark (XEB), which correlates sampled outcomes with ideal probabilities, is a fidelity proxy only under conventional PT statistics~\cite{Boixo_2018,Arute_2019,Dalzell_2024,Morvan_2024,ware2023sharpphasetransitionlinear}. Free-fermion circuits, whose ideal probabilities are efficiently computable, would be a natural testbed for such benchmarks~\cite{Helsen_2022}, but only once their own output statistics are known.

Here we tackle this question by analytically deriving the exact statistics of the measurement probabilities of random FGS, both with and without particle number conservation. We show that the logarithm of the rescaled probabilities becomes Gaussian at large system size $L$, and that random FGS anticoncentrate only up to polynomial factors in $L$. We then show how this fermionic analogue of the Porter--Thomas law emerges in random brickwork matchgate circuits in $1+1$D. To this end, we first restrict ourselves to Clifford matchgates, which allows us to map the problem exactly onto a simple model of arcs and loops. We show that convergence to the fermionic PT law is characterized by an early diffusive regime, with deviations $\propto L/\sqrt t$, followed by slowest-mode relaxation $e^{-\pi^2t/L^2}$. We use this result to rigorously prove that one-dimensional matchgate circuits require a depth $t\gtrsim L^2/\log^2L$ for the collision probability to converge to its asymptotic value. Finally, we show that ideal free-fermion sampling yields a linear cross-entropy score growing as $\sqrt L$, and we propose a log-based benchmark with an exact fermionic normalization.

\section{Setup}
We consider $L$ fermionic modes, with operators $c_j,c_j^\dagger$, $j\in\{1,\dots,L\}$. The Majorana operators are $\gamma_{2j-1}=c_j+c_j^\dagger$ and $\gamma_{2j}=i(c_j^\dagger-c_j)$. The computational basis corresponds to the Fock states $|x\rangle=(c_1^\dagger)^{x_1}\cdots(c_L^\dagger)^{x_L}|0\rangle$, with $x_j\in\{0,1\}$ the local occupation
number and $|0\rangle=|0\dots0\rangle$ the Fermi vacuum. Fermionic Gaussian unitaries (FGUs), or matchgates, are unitaries whose action by conjugation on the Majoranas is an orthogonal transformation $Q$, i.e.\ $U\gamma_\mu U^\dagger=\sum_{\nu=1}^{2L}Q_{\mu\nu}\gamma_\nu$~\cite{Surace_2022}. For $Q\in SO(2L)$, writing $Q=e^K$ with $K^T=-K$ gives, up to a phase, $U_Q=\exp\!\big(\tfrac14\sum_{\mu,\nu}K_{\mu\nu}\gamma_\mu\gamma_\nu\big)$; so these unitaries are generated by quadratic Majorana operators and commute with the parity $P=\prod_j Z_j$, with $Z_j=-i\gamma_{2j-1}\gamma_{2j}$. A pure \emph{fermionic Gaussian state (FGS)} is $|\psi\rangle=U_Q|0\rangle$; it has definite parity, but generally not a fixed particle number. A random FGS is obtained by drawing $Q$ from the Haar measure on $SO(2L)$. The number-preserving FGUs, which commute with the particle number $N=\sum_{j=1}^Lc_j^\dagger c_j$, form a subgroup isomorphic to $U(L)$. A pure $N$-particle number-preserving FGS is $|\psi\rangle=U_V|1^N0^{L-N}\rangle=\prod_{\alpha=1}^N\big(\sum_{j=1}^LV_{j\alpha}c_j^\dagger\big)|0\rangle$, and a random one is obtained by drawing $V$ from the Haar measure on $U(L)$. Finally, Clifford matchgates are the FGUs that are also Clifford unitaries, i.e.\ they map each Majorana operator to a single Majorana operator up to a sign, $U\gamma_\mu U^\dagger=s_\mu\gamma_{\sigma(\mu)}$ with $\sigma\in S_{2L}$ and $s_\mu=\pm1$, so that $Q_{\mu\nu}=s_\mu\delta_{\nu,\sigma(\mu)}$ is a signed permutation matrix (with $\det Q=1$). A random \emph{Clifford FGS} is $|\psi\rangle=U_Q|0\rangle$, with $Q$ drawn uniformly among the signed permutation matrices with $\det Q=+1$. For each of these three ensembles of random states, generically denoted by $\mathcal{E}$, we study the statistics of basis-state probabilities $p_{\psi}(x)=|\langle x|\psi\rangle|^2$, defined as
\begin{equation}
P(w)=\mathbb{E}_{\psi\sim\mathcal{E}}\left[D^{-1} \sum_{x\in\mathcal{X}}\delta\bigl(w-D \, p_{\psi}(x)\bigr) \right] \, .
\end{equation}
Here, $\mathcal{X}$ is the set of allowed bitstrings and $D=|\mathcal{X}|$. Specifically, $D=2^{L-1}$ for FGS in the even (or odd) parity sector, and $D=\binom{L}{N}$ in the particle-number-conserving case. The moments $\E[w^k]=D^{k-1}\sum_{x\in\mathcal X}\E_{\psi\sim\mathcal E}[p_\psi(x)^k]$ are proportional to the $k$-th inverse participation ratios~\cite{Turkeshi_2024_1,Misguich_2016,Sierant_2022,Sierant_2025}. By invariance of the ensembles, they also coincide, up to a factor $D^k$, with the frame potentials $\E_{\psi,\varphi\sim\mathcal E}[|\langle\psi|\varphi\rangle|^{2k}]$, computed for fermionic Gaussian ensembles in Refs.~\cite{tiutiakina2023framepotentialbrowniansyk,sierant2026theorymatchgatecommutant}. For Haar-random states, $w$ follows the PT distribution $P_{\rm PT}(w)\overset{D\gg1}{\simeq} e^{-w}$, with $\E[w^k]\overset{D\gg1}{\simeq} k!$.

% =\frac{D-1}{D}\big(1-\frac wD\big)^{D-2}
\section{PT statistics for random FGS}

\begin{figure}[ht!]
    \centering
    \includegraphics[width=1.01\linewidth]{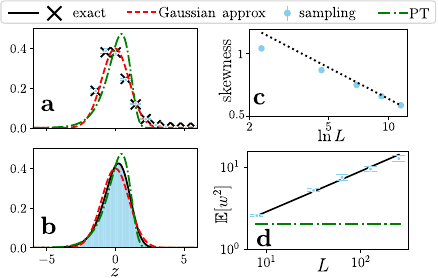}
     \caption{\textbf{Porter--Thomas statistics for random FGS.} (a,b) Distribution of the standardized log-probabilities $z=(\ln w-\E[\ln w])/\sqrt{\mathrm{Var}[\ln w]}$: exact result (crosses, black line) and sampling (dots, histogram). (a) Clifford FGS, restricted to nonzero probabilities ($L=2000$). (b) Generic non-number-preserving FGS ($L=32$). At finite $L$ both differ from a Gaussian (red), and from conventional Porter--Thomas (green). (c) Skewness of $z$ for Clifford FGS, decaying as $(\ln L)^{-1/2}$ (dotted line). (d) Second moment $\E[w^2]$ for generic FGS: sampling (dots) and exact result (black line), which grows as $\sim \sqrt{L}$, whereas $\E[w^2]=2$ for Porter--Thomas (green line). Random FGS thus anticoncentrate only up to polynomial factors.}
    \label{fig:plots1}
\end{figure}

\paragraph{Number-preserving FGSs} -- We set $|\psi\rangle=U_V|x_0\rangle$, with Haar-distributed $V\in U(L)$ and $N=|x_0|$ number of particles. We write $S_x$ for the occupied set of $x$, and $V[A,B]$ for the submatrix of $V$ with rows $A$ and columns $B$. The bitstring probabilities are given by the Slater determinant $p_{\psi}(x) = |\langle x|U_V|x_0\rangle|^2=|\det V[S_x,S_{x_0}]|^2$. Permutation matrices leave the Haar measure invariant, so we can always reduce to study the distribution of $w \eqd D |\det T|^2$ with $T=V[[N],[N]]$ the top $N \times N$ minor. Here and in the following we use $\eqd$ to denote equality in probability distribution. 

\begin{prop}[1]
For random number-preserving FGS, the probabilities are distributed as the product of $N$ independent $\Betad$-distributed random variables. Specifically, $w \eqd D \prod_{j=1}^N B_j$, with $B_j\sim\Betad(j,L-N)$. Moreover, the associated $k-$th moment is $\E[w^k]= D^k \prod_{r=0}^{k-1}\frac{(N+r)!\,(L-N+r)!}{r!\,(L+r)!}$~\cite{Collura_2026,sierant2026theorymatchgatecommutant}
\end{prop}

The proof is elementary and uses only a few standard properties of the Haar measure. Beyond the moments, the product structure gives directly the full distribution of $w$ (as the density of $\ln w$ is the convolution of those of the independent $\ln B_j$). This structure reflects the free-fermion nature of FGS: each particle contributes an independent factor, restricted by Pauli exclusion to the space left free by the others. Due to this structure, it is natural to consider $\ln w=\ln D+\sum_{j=1}^N\ln B_j$, a sum of independent terms. Its mean and variance follow from $\E[\ln B]=\psi(a)-\psi(a+b)$ and $\Var[\ln B]=\psi'(a)-\psi'(a+b)$ for $B\sim\Betad(a,b)$, with $\psi$ the digamma function:
\begin{equation}\label{eq:expansion1}
\E[\ln w]=-\tfrac12\ln\Lambda+O(1),\qquad \Var[\ln w]=\ln\Lambda+O(1),
\end{equation}
at large $L$ and fixed density $\nu=N/L$, with $\Lambda=\nu(1-\nu)L$. The variance diverges, while the higher cumulants of each term decay fast enough that their sums stay bounded. By the central limit theorem (Lyapunov formulation), $\ln w$ is therefore asymptotically Gaussian, with mean and variance of order $\ln L$. The convergence is slow, with skewness $O((\ln L)^{-3/2})$. Equivalently, $w$ is asymptotically log-normal. 

\paragraph{Generic FGSs} -- We set $|\psi\rangle=U_Q|0\rangle$, with $Q$ drawn from the Haar measure on $SO(2L)$. The probabilities $p_\psi(x)=|\langle x|U_Q|0\rangle|^2$ are no longer Slater determinants, but follow from the general overlap formula between two FGS in terms of their covariance matrices~\cite{Mazza_2012}. Since $U_Q$ preserves the fermion parity and the vacuum is even, only even bitstrings $x$ have nonzero probability; we thus set $D=2^{L-1}$ and $w=D\,p_\psi(x)$. We find a structure similar to before.

\begin{prop}[2]
For random FGS, $w \eqd D \prod_{j=1}^{L-1} B_j$, with independent $B_j\sim\Betad(j,j)$. The moments are
$\E[w^k]= D^k \prod_{1\le i<j\le L}\frac{2L-i-j}{k+2L-i-j}$.
\end{prop}

As before, the proof we provide is elementary. Similarly, $\E[\ln w]=\ln D+\sum_{j=1}^{L-1}\big[\psi(j)-\psi(2j)\big]$ and
$\mathrm{Var}[\ln w]=\sum_{j=1}^{L-1}\big[\psi'(j)-\psi'(2j)\big]$, which at large $L$ expand as
\begin{align}\label{eq:expansion2}
\begin{split}
\E[\ln w] = -\tfrac{\ln L}{4} +O(1) \qquad \mathrm{Var}[\ln w] =\tfrac{\ln L}{2}+O(1) \,.
\end{split}
\end{align}
Hence, the distribution of $w$ is asymptotically log-normal, with half the log-variance of the number-preserving case (see Figure~\ref{fig:plots1}b). 

\paragraph{Clifford FGSs} -- We set $|\psi\rangle=U_Q|0\rangle$ with $Q$ a random signed permutation matrix, $U_Q\gamma_\mu U_Q^\dagger=s_\mu\gamma_{\sigma(\mu)}$. Since $Z_j=-i\gamma_{2j-1}\gamma_{2j}$ stabilize the vacuum, $|\psi\rangle$ is stabilized by the $L$ operators $-i\,s_{2j-1}s_{2j}\,\gamma_{\sigma(2j-1)}\gamma_{\sigma(2j)}$. The pairs $\{\sigma(2j-1),\sigma(2j)\}$ form a uniformly random perfect matching of the $2L$ Majoranas~\cite{w97w-7zny}. Drawing the Majoranas as dots, the matching is a set of $L$ arcs, each joining a pair of dots~\cite{w97w-7zny}. Our first result shows that basis-state probabilities can be read directly from this arc picture: whenever nonzero, $p_\psi(x)=2^{-L+c}$, where $c$ is the number of loops formed by the $L$ arcs of $|\psi\rangle$ together with the $L$ arcs $(2j-1,2j)$ associated to $|x \rangle$. Whether $p_\psi(x)$ is nonzero is decided by a parity condition. Indeed, any diagonal stabilizer operator $g=\pm\{\Id,\PauliZ\}^{\otimes L}$ of $|\psi\rangle$ gives $\langle x|\psi\rangle=\langle x|g|\psi\rangle=g(x)\langle x|\psi\rangle$, so $\langle x|\psi\rangle\neq0$ requires $g(x)=+1$. Graphically, we colour each arc $\varepsilon(-i\gamma_a\gamma_b)$ ($a<b$) blue if $\varepsilon=+1$ and red if $\varepsilon=-1$. Then $p_\psi(x)\neq0$ if on every loop, the number of red arcs plus the number of crossings between arcs of $|\psi\rangle$ is even. For example, consider the state $(|00\rangle + i |11\rangle) / \sqrt{2}$ (stabilized by $XY = -i \gamma_2 \gamma_4$ and $YX =+ i \gamma_1 \gamma_3$). The overlaps (squared) with $00$ and $01$ are respectively 
\tikzset{dot/.style={circle,fill,inner sep=1.3pt},
         plus/.style={thick,blue!70!black}, minus/.style={thick,red!80!black}}
%\begin{equation}
%\begin{tikzpicture}[baseline=-0.5ex]
%  \foreach \i in {1,...,4} \node[dot] (g\i) at (0.7*\i-0.7,0) {};
%  \draw[minus] (g1) to[out=-90,in=-90,looseness=1.3] (g3);        % state arc  YX = +i g1 g3  (eps = -1)
%  \draw[plus]  (g2) to[out=-90,in=-90,looseness=1.3] (g4);        % state arc  XY = -i g2 g4  (eps = +1)
%\end{tikzpicture}
%\end{equation}
\begin{equation}
2^{-1}  \leftrightarrow
\begin{tikzpicture}[baseline=-1.5ex]
  \foreach \i in {1,...,4} \node[dot] (g\i) at (0.7*\i-0.7,0) {};
  \draw[plus]  (g1) to[out=90,in=90,looseness=1.6] (g2);          % basis arc, x_1 = 0
  \draw[plus]  (g3) to[out=90,in=90,looseness=1.6] (g4);          % basis arc, x_2 = 0
  \draw[minus] (g1) to[out=-90,in=-90,looseness=1.3] (g3);        % state arc  YX = +i g1 g3  (eps = -1)
  \draw[plus]  (g2) to[out=-90,in=-90,looseness=1.3] (g4);        % state arc  XY = -i g2 g4  (eps = +1)
  %\node at (1.05,-1.0) {\small $x=00$: $p_x=\tfrac12$};
\end{tikzpicture}
\, \, \text{   and    } \, \, 
0 \leftrightarrow
\begin{tikzpicture}[baseline=-1.5ex]
  \foreach \i in {1,...,4} \node[dot] (g\i) at (0.7*\i-0.7,0) {};
  \draw[plus]  (g1) to[out=90,in=90,looseness=1.6] (g2);          % x_1 = 0
  \draw[minus] (g3) to[out=90,in=90,looseness=1.6] (g4);          % x_2 = 1
  \draw[minus] (g1) to[out=-90,in=-90,looseness=1.3] (g3);
  \draw[plus]  (g2) to[out=-90,in=-90,looseness=1.3] (g4);
  %\node at (1.05,-1.0) {\small $x=01$: $p_x=0$};
\end{tikzpicture}
\end{equation}
Since each loop gives one parity condition, a state with $c$ loops has nonzero overlap with $2^{L-c}$ basis states, i.e.\ a fraction $2^{1-c}$ of the even bitstrings, all with $w=2^{c-1}$. The distribution of $w$ thus reduces to that of the number of loops $c$ of a uniformly random perfect matching, which by simple counting is $P_c(c)=2^{L-c}\,|s(L,c)|/(2L-1)!!$, where $|s(L,c)|$ is the number of permutations of $L$ elements with $c$ cycles. Then, $u\equiv(\log_2 w\,|\,w\neq0)=c-1$ is distributed as $P(u)\propto 2^{-(u+1)}P_c(u+1)$. We therefore arrive at

\begin{prop}[3]
For random Clifford FGS, a fraction $P(w\neq0)=\E[2^{1-c}]=2\prod_{j=0}^{L-1}\frac{4j+1}{4j+2}$ of the
probabilities is nonzero. On the nonzero probabilities, $\log_2 w$ is distributed as a sum of $L-1$ independent Bernoulli
variables: $u\eqd\sum_{j=1}^{L-1}B_j$ with $B_j\sim\mathrm{Bernoulli}\big(\frac1{4j+1}\big)$. The distribution is $P(u)=4^{L-u-1}\,|s(L,u+1)| \, \prod_{j=0}^{L-1}(4j+1)^{-1}$.
Moreover, the moments of $w$ are $\mathbb E[w^k]=2^{1-k}\prod_{j=1}^{L}\left(1+\frac{2^{k-1}-1}{2j-1}\right)$.
\end{prop}

For $k\leq3$ the moments of $w$ coincide with those of random (non-number-preserving) FGS obtained in Proposition~2, consistently with the fact that Clifford matchgates form a 3-design for the group of fermionic Gaussian unitaries~\cite{Wan_2023,Braccia_2025,sierant2026theorymatchgatecommutant}, i.e.\ they reproduce all its Haar averages up to three copies of the state. Since $u$ is a sum of independent Bernoulli variables, its mean and variance follow directly, $\mathbb E[u]=\sum_{j=1}^{L-1}\frac{1}{4j+1} \overset{L \gg 1}{=} \frac14\ln L+O(1)$ and $\operatorname{Var}[u]=\sum_{j=1}^{L-1}\frac{4j}{(4j+1)^2} \overset{L \gg 1}{=} \frac14\ln L+O(1)$. As the variance diverges, the central limit theorem implies that $u$, too, becomes Gaussian, i.e.\ $w$ is again log-normal. The convergence is however slower than for generic FGS, with skewness $\simeq(\frac14\ln L)^{-1/2}$, since all cumulants of $u$ grow as $\frac14\ln L$.

\paragraph{Consequences for anticoncentration} -- From the lognormal moments and the expansions~\eqref{eq:expansion1} and~\eqref{eq:expansion2}, we get $\E[w^k]\sim\Lambda^{k(k-1)/2}$ for number-preserving FGS and $\E[w^k]\sim L^{k(k-1)/4}$ for generic FGS, while Proposition~3 gives $\E[w^k]\sim L^{(2^{k-1}-1)/2}$ for Clifford FGS. In all cases the moments grow polynomially with $L$, whereas for conventional PT $\E[w^k]=k!$ stays $O(1)$. Random FGS thus do not anticoncentrate in conventional sense, but only up to polynomial factors (see Figure~\ref{fig:plots1}d). 

\section{Random matchgate circuits}

We now study how the distribution of $w$, derived above for globally random states, emerges in random brickwork matchgate circuits in $1+1$D, with alternating even and odd layers, and with qubits initialized in the vacuum state $|0 \rangle$. We first treat the Clifford case, where the arc picture allows exact analytics, and then extend the analysis to generic matchgates.

\begin{figure*}[ht!]
    \centering
    \includegraphics[width=1.02\linewidth]{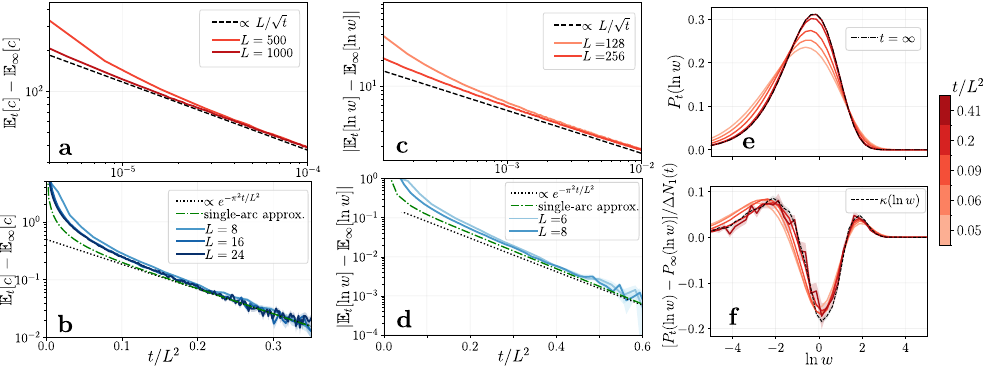}
     \caption{\textbf{Emergence of fermionic Porter--Thomas statistics in random $1+1$D matchgate circuits.} (a,b) Clifford circuits: deviation of the average number of loops $\E_t[c]$ from its asymptotic value $\E_\infty[c]$, versus rescaled depth $t/L^2$. (a) Early diffusive regime, $\E_t[c]-\E_\infty[c]\propto L/\sqrt t$. (b) Late exponential relaxation $\propto \exp(-\pi^2t/L^2)$, well captured by the single-arc approximation (dash-dotted for $L=24$). (c,d) Generic random matchgate circuits: the deviation $|\E_t[\ln w]-\E_\infty[\ln w]|$ displays the same two regimes. (e) Full distribution $P_t(\ln w)$ approaching its asymptotic form ($L=8$, $2\cdot10^5$ circuit realizations). (f) Its finite-depth deviations, rescaled by $\Delta N_1(t)=\E_t[N_1]-\E_\infty[N_1]$, collapse onto the single-arc prediction $\kappa(\ln w)$ (dashed, estimated by sampling Haar-random FGS).}
    \label{fig:plots2}
\end{figure*}

\paragraph{Random Clifford matchgate circuit} -- The initial state is the matching $\pi = \{(2j-1,2j)\}_{j=1}^L$. Each Clifford matchgate permutes the four Majoranas of its two sites uniformly at random (up to signs), thus moving the endpoints of the arcs, as illustrated here:
\tikzset{dot/.style={circle,fill,inner sep=1.3pt},
         arc/.style={thick},
         wire/.style={gray,thin,-{Latex[length=2.5pt]}}}   % needs \usetikzlibrary{arrows.meta}
\begin{equation}
\begin{tikzpicture}[baseline=(current bounding box.center)]
  \foreach \i in {1,...,4} {\node[dot] (b\i) at (0.7*\i-0.7,0) {}; \node[dot] (t\i) at (0.7*\i-0.7,0.7) {};}
  \draw[arc] (b1) to[out=-90,in=-90,looseness=0.9] (b2);   % initial arcs (1,2), (3,4)
  \draw[arc] (b3) to[out=-90,in=-90,looseness=0.9] (b4);
  \draw[arc] (t1) to[out=90,in=90,looseness=0.9] (t3);     % arcs after the permutation: (1,3), (2,4)
  \draw[arc] (t2) to[out=90,in=90,looseness=0.9] (t4);
  \begin{scope}[shorten <=2.5pt, shorten >=2.5pt]          % gap between arrows and dots
    \draw[wire] (b1) -- (t3);                              % permutation 1->3, 2->1, 3->4, 4->2
    \draw[wire] (b2) -- (t1);
    \draw[wire] (b3) -- (t4);
    \draw[wire] (b4) -- (t2);
  \end{scope}
\end{tikzpicture}
\end{equation}
The overall process is therefore a Markov chain on the $(2L-1)!!$ perfect matchings, $P_{t+1}(\pi')=\sum_{\pi}M^{(t\bmod 2)}_{\pi'\pi}\,P_t(\pi)$ where $t\bmod2$ is the parity of the layer and $M^{(0)},M^{(1)}$ are $(2L-1)!!\times(2L-1)!!$ transition matrices. Since a permutation and its inverse are equally likely, both matrices are symmetric and doubly stochastic. Moreover, the chain is irreducible (any matching can be reached from any other) and aperiodic (each gate leaves the matching unchanged with nonzero probability), so by the Perron--Frobenius theorem $P_\infty(\pi)=1/(2L-1)!!$ is the unique stationary state. 
Expanding in the eigenvectors $V_\lambda$ of the two-layer transfer matrix $M=M^{(1)}M^{(0)}$, with eigenvalues $|\lambda|<1$, gives $P_t=M^{t/2}P_0=P_\infty+\sum_\lambda c_\lambda\lambda^{t/2}V_\lambda$.
Given the complexity of the full perfect matching process, one can instead focus \emph{on the evolution of a single arc}. This is described by a second Markov chain $P_{t+1}(p')=\sum_{p} K^{(t\bmod 2)}_{p'p}\,P_t(p)$ in the $\binom{2L}{2}$-dimensional space of pairs $p=(a,b)$ (with $a,b \in \{1, \dots, 2L\}$). This chain is a variant of the symmetric exclusion process~\cite{SPITZER1970246,Liggett1999StochasticIS}. Interestingly, an exact duality between the two processes holds~\cite{Liggett2005}. In fact, first notice that, for any vector $f_p$ in the pair space, we can associate a vector in the matching space as $V_f(\pi)=\sum_p f_p \, n_{p}(\pi)$, where $n_{p}(\pi)$ is an indicator function, equal to $1$ if the arc $p$ is present in $\pi$ and $0$ otherwise. Second, one can easily prove (see End matter) that $M^{(\alpha)}\,V_f=V_{K^{(\alpha)}f}$ for each layer $\alpha\in\{0,1\}$, i.e.\ on vectors of this form the action of the evolution on the matchings reduces exactly to that of the single-arc chain on the pairs. This exact equality follows from the fact that, although the arcs evolve in a correlated way (two arcs cannot share a dot), $V_f$ is linear in the arcs, and each arc separately evolves according to the single-arc chain. As a consequence, every eigenvector $f$ of the single-arc chain over one period, $K f=\lambda f$ (with $K=K^{(1)}K^{(0)}$) can be lifted to an exact eigenvector of the full chain with the same eigenvalue, $M V_f=\lambda V_f$. To gain a better understanding of the overall matching process, we thus approximate it by its projection onto the `single-arc subspace' $\mathcal S=\mathrm{span}\{V_f\}$, spanned by the vectors $V_f$ built from arcs. To this end, we replace each layer $M$ by $QMQ$, with $Q$ the orthogonal projector onto $\mathcal S$. $Q$ commutes with $M$ and the dynamics within $\mathcal S$ is treated exactly (see End Matter). The approximation then consists only in discarding the component of the initial state outside $\mathcal S$, $P_t\simeq QP_t=M^tQP_0$.
The spectrum of $QMQ$ consists of the lifted single-arc modes. However, two kinds of modes drop out: i) those of a single Majorana, $g(a,b)=h_a+h_b$, for which $V_g=0$; ii) for even $L$, those that are odd under the reflection of the chain, since they have vanishing overlap with the initial vacuum state. We now distinguish two time regimes. \\
At late times, only the slowest mode survives. Its form can be understood by approximating the two endpoints of the arc as independent random walkers. A single walker on sites $j\in\{1,\dots,L\}$ has eigenmodes $h_n(j)\simeq\cos\big(n\pi(j-\tfrac12)/L\big)$ and eigenvalues $\cos^2(n\pi/L)$. Products $h_m(a)h_n(b)$ are then approximate eigenvectors of the arc process with eigenvalue $\cos^2 \frac{m \pi}{L} \cos^2 \frac{n \pi}{L}$. Those with $n=0$ are single-Majorana modes, which disappear. The slowest mode is therefore $f_{a,b}\simeq h_1(j_a)\,h_1(j_b)$, with $j_a$ the site of dot $a$, and eigenvalue $\lambda\simeq\cos^4\frac{\pi}{L}$. Figure~\ref{fig:plots2} shows the average number of loops $\E_t[c]=\sum_\pi c(\pi)P_t(\pi)$, decreasing from $L$ to $\E_\infty[c]=\sum_{j=1}^L\frac{1}{2j-1}\simeq\frac12\ln L$. At late times, the decay $\lambda^{t/2}\simeq e^{-\pi^2t/L^2}$ (dotted line) accurately captures the exact dynamics, a nontrivial check since slower modes could in principle exist outside the single-arc sector. \\
Beyond the slowest mode, the single-arc projection can be solved in closed form. For any site-invariant observable $V(\pi)$, one finds $QV=\E_\infty[V]+\kappa_V\big(N_1-\E_\infty[N_1]\big)$, where $N_1(\pi)=\sum_jn_{(2j-1,2j)}(\pi)$ counts the single-site loops and $\kappa_V$ is a normalized overlap between $V$ and $N_1$, computed over uniformly random matchings (see End Matter). Thus
\begin{equation}\label{eq:projectionV}
\E_t[V]-\E_\infty[V]\simeq\kappa_V\big(\E_t[N_1]-\E_\infty[N_1]\big),
\end{equation}
where $\E_t[N_1]$ follows exactly from the single-arc chain, since $N_1\in\mathcal S$. All site-invariant observables therefore relax with the same time dependence and differ only through $\kappa_V$. At late times, this reproduces the slowest mode. At early times the two ends of an arc diffuse and $N_1$ counts the arcs whose ends are back on the same site, so $\E_t[N_1]\simeq L/(4\sqrt{\pi t})$ for $1\ll t\ll L^2$, i.e.\ an extensive, diffusive relaxation $\propto L/\sqrt t$. Figure~\ref{fig:plots2} shows the case of the number of loops, for which $\kappa_c=\frac{2L-3}{2L-2}$. At late times the projection reproduces the slowest-mode decay. At early times it gives the correct $L/\sqrt t$ scaling (and since $c\ge N_1$, $\E_t[c]\ge\E_t[N_1]$ is a rigorous bound). However, it underestimates the prefactor: it predicts $\E_t[c]-\E_\infty[c]\simeq L/(4\sqrt{\pi t})\approx0.14\,L/\sqrt t$, while numerically $\E_t[c]-\E_\infty[c]\simeq0.37\,L/\sqrt t$. \\

\paragraph{Generic random matchgate circuit} -- The arc picture extends to generic matchgate circuits. For a Clifford FGS, $\Gamma_p^2=1$ if $p=(a,b)$ is an arc and $0$ otherwise, so that the arc indicators are $n_p(\pi)=\Gamma_p^2$; for a generic FGS, the same quantities $\Gamma_p^2$ take values in $[0,1]$. Since Clifford matchgates form a 3-design~\cite{Wan_2023,Braccia_2025}, averaging over a Haar-random gate acts on the quadratic quantities $\Gamma_p^2$ exactly as averaging over a Clifford gate, i.e.\ as the single-arc chain $K$. Hence $\E_t[N_1]$, with $N_1=\sum_j\Gamma_{2j-1,2j}^2=\sum_j\langle Z_j\rangle^2$, is the same as in the Clifford case, and the single-arc approximation takes the same form as in Eq.~\eqref{eq:projectionV}, now with $\kappa_V$ computed over Haar-random FGS rather than uniform matchings.

Figure~\ref{fig:plots2}(c,d) shows that $\E_t[\ln w]$ relaxes as in the Clifford case, first diffusively as $L/\sqrt t$, then as $e^{-\pi^2t/L^2}$. The same formula, applied to each bin of the histogram, describes the full distribution: as $P_t(\ln w)$ approaches the fermionic Porter--Thomas law (e), all its deviations collapse onto a single, time-independent profile $\kappa(\ln w)$ (f). The single-arc projection thus provides an excellent approximation of the whole distribution $P_t(\ln w)$, already from depths $t\gtrsim0.05\,L^2$, well before the slowest mode dominates.

These results indicate that matchgate circuits need a depth $t\sim L^2$ to reach the fermionic Porter--Thomas law. This can be made rigorous for the collision probability $Z_t\equiv\E_t[w^2]$. We say that a circuit anticoncentrates (in the fermionic sense) once $Z_t\le q \, Z_\infty$ for a fixed $q>1$. Combining the exact dynamics of $\E_t[N_1]$ with $c\ge N_1$, we prove (see End Matter):
\begin{theorem}
There is a constant $C$ such that, for both generic and Clifford matchgate brickwork circuits in $1+1$D and any $q>1$,
$Z_t\le q\,Z_\infty\ \Longrightarrow\ t\ge L^2/\big(C\log_2^2(qL)\big)$.
\end{theorem}

\section{Conclusions}
We have derived the fermionic analogue of the Porter--Thomas law: probabilities of random FGS are asymptotically log-normal, with moments growing polynomially in system size. We also showed that random matchgate circuits approach this law only at depths scaling as $L^2$. A direct consequence concerns the linear cross-entropy benchmark $\text{XEB}=D\,\E_{x\sim\tilde p}[p_x]-1$, commonly used to certify sampling experiments, with $\tilde p$ the actual sampled distribution. For noiseless sampling ($\tilde p=p$) it equals $\E[w^2]-1$, which is $1$ for Porter--Thomas but grows as $\sqrt L$ for random FGS. Free-fermion circuits, though classically simulable, thus reach large XEB scores. We therefore propose the log-based score $\text{logXEB}=\E_{x\sim\tilde p}[\ln w_x]$~\cite{Boixo_2018}, which is better suited since $\ln w$ is nearly Gaussian. Using Proposition~2, for random FGS we find $\text{logXEB}=(L-1)\ln2+\sum_{j=1}^{L-1}[\psi(j+1)-\psi(2j+1)]$ for noiseless sampling and $\text{logXEB}=(L-1)\ln2+\sum_{j=1}^{L-1}[\psi(j)-\psi(2j)]$ for uniform sampling, $\tilde p=1/D$. The gap $\Delta=\frac12\sum_{j=1}^{L-1}j^{-1}\simeq\frac12\ln L$ can be used to normalize the score $F=(\text{logXEB}-\text{logXEB}_{\rm unif})/\Delta$, which then interpolates between $0$ and $1$. It remains to be studied whether $F$ provides a faithful estimate of the fidelity under realistic noise models, beyond the global depolarizing case where it is exact. 

\section{Acknowledgements}
I thank Arman Sauliere, Andrea De Luca, Jacopo De Nardis for our fruitful collaborations on topics related to this article, and I thank Xhek Turkheshi, Piotr Sierant, Alessio Paviglianiti for discussions and useful feedback on the problems addressed in this paper. Finally, I am deeply grateful to G.M. for her love and our intellectually stimulating discussions. Part of this work was conceived and carried out during a workshop at ECT* (Trento), whose hospitality is gratefully acknowledged.

\bibliography{bib}
\clearpage
\appendix 

\section{End Matter}

\paragraph{Proof of proposition (1)} -- We recall that for a uniform random unit vector $v\in\mathbb C^L$, the weight on $N$ components is $\Betad$-distributed: $\left(\sum_{i=1}^N|v_i|^2 \right) \eqd \Betad(N,L-N)$. %For $N=1$, this observation reduces to the conventional Porter--Thomas law.

\begin{proof}
Let $v_1,\dots,v_N$ be the first $N$ columns of $V$ (so $T$ is their top $N\times N$ block), and $\mathcal P(L,N)$ the distribution of $|\det T|^2$. Multiplying $V$ on the left by $A\oplus B$, with $A\in U(N)$ acting on the top $N$ coordinates and $B\in U(L-N)$ on the bottom $L-N$, changes $\det T$ only by the phase $\det A$, even if $A,B$ depend on $V$. Choosing them according to $v_1$, we may take $v_1=c\,e_1+\sqrt{1-c^2}\,e_{N+1}$ with $c\ge0$ and $c^2\sim\Betad(N,L-N)$ by the fact above. The first column of $T$ is then $(c,0,\dots,0)^T$, so $|\det T|=c\,|\det T'|$, with $T'$ the block of rows and columns $2,\dots,N$. Conditional on $v_1$, the columns $v_2,\dots,v_N$ are Haar-random orthonormal vectors in $v_1^\perp$, of dimension $L-1$, so $|\det T'|^2\sim\mathcal P(L-1,N-1)$ independently of $c$, and $\mathcal P(L,N)=\Betad(N,L-N)\times\mathcal P(L-1,N-1)$. Iterating down to $\mathcal P(L-N,0)=\delta_1$ gives $w\eqd D\,\Betad(N,L-N)\cdots\Betad(1,L-N)$. Finally, $\E[B^k]=\prod_{r=0}^{k-1}\frac{a+r}{a+b+r}$ for $B\sim\Betad(a,b)$ gives $\E[w^k]=D^k\prod_{r=0}^{k-1}\prod_{j=1}^N\frac{j+r}{L-N+j+r}$, where the $j$-product telescopes into the factorials.
\end{proof}

\paragraph{Proof of proposition (2)} -- We use the covariance matrix: $\Gamma_{ab}=i\langle\psi|\gamma_a\gamma_b|\psi\rangle$ ($a\neq b$), $\Gamma_{aa}=0$~\cite{Surace_2022}. It is real and antisymmetric, and for a pure Gaussian state it is also orthogonal, $\Gamma^T \Gamma= \Id$. For the vacuum state, it reads $(\Gamma_0)_{2j-1,2j}=-1$. For $|\psi\rangle=U_Q|0\rangle$, one has $\Gamma=Q\Gamma_0Q^T$. By Haar invariance, all even bitstrings are equivalent, so we consider $p_\psi(0)$. 

\begin{proof}
As in the number-preserving case, we proceed iteratively, removing one mode (instead of one particle) per iteration. The proof has three stages. \emph{Step 1.} Let $\Pi=(\Id+\PauliZ_L) / 2$ be the projector onto $|0\rangle_L$, $q = \langle\psi|\Pi|\psi\rangle$ the probability that mode $L$ is empty, $|\psi'\rangle :=(\langle 0|_L)|\psi\rangle /  \, \sqrt q$ the resulting state of modes $1,\dots,L-1$, and $p_0' = | \langle 0 \dots 0|\psi'\rangle|^2$. Notice that $|\psi'\rangle$ is Gaussian, as $\Pi=\lim_{\beta\to\infty}e^{-\beta c_L^\dagger c_L}$ is a limit of exponentials of quadratic operators~\cite{Coppola_2022}. Moreover, it is even. \emph{Step 2.} We have $q=(1+\langle \PauliZ_L\rangle) / 2=(1-\Gamma_{2L-1,2L})/2$, so we need the law of one entry of $\Gamma$. Consider the last column $v=\Gamma e_{2L}$ of $\Gamma$. It is a unit vector, because $\Gamma$ is orthogonal, and it is orthogonal to $e_{2L}$, because $\Gamma$ is antisymmetric. So $v$ lies on the unit sphere $S^{2L-2}$ of the hyperplane $e_{2L}^\perp$. We want to show that $v$ is distributed uniformly on this sphere. To see this, consider a rotation $R\in SO(2L)$ that fix $e_{2L}$. Because of Haar invariance, $\Gamma'=R\Gamma R^T$ is distributed as $\Gamma$. Moreover $v' = \Gamma'e_{2L}=Rv$, which implies $Rv\eqd v$. So the law of $v$ is the uniform probability measure on $S^{2L-2}$, the only rotation-invariant one. Now notice that $\Gamma_{2L-1,2L}=\langle e_{2L-1},v\rangle$ is one of the coordinates of $v$. One coordinate $t$ of a uniform unit vector in $\mathbb R^d$ has density $\propto(1-t^2)^{(d-3)/2}$. With $d=2L-1$, this gives $q\sim\Betad(L-1,L-1)$. \emph{Step 3.} Let $U_R$, $R\in SO(2L-2)$, act only on $\gamma_1,\dots,\gamma_{2L-2}$. It leaves $q$ unchanged and maps $|\psi'\rangle$ to $U_R|\psi'\rangle$, so by Haar invariance $(q,|\psi'\rangle)\eqd(q,U_R|\psi'\rangle)$. Hence, at fixed $q$, $|\psi'\rangle$ is Haar-distributed on $L-1$ modes; since this does not depend on $q$, $|\psi'\rangle$ is independent of $q$, and $p_0'$ has the law of $p_0$ for $L-1$ modes. \emph{Conclusion.} $p_0(L)\eqd\Betad(L-1,L-1)\times p_0(L-1)$ with independent factors, and iterating down to $L=1$ proves the proposition.
\end{proof}

\paragraph{Proof of Proposition (3)} -- First, we derive the distribution of the number of loops $c$ of a random perfect matching. Going around a loop, one visits a set of sites, each site $j$ carrying two dots (Majorana modes) $2j-1,2j$. A loop through $r$ given sites is fixed by a directed cyclic order, $(r-1)!$ choices, and an entry dot at each site, $2^r$ choices; reversing the direction gives the same loop with all entry dots swapped, so there are $(r-1)!\,2^{r-1}$ such loops. A matching with $c$ loops is fixed by a partition of the $L$ sites into $c$ blocks and a loop through each block, so
\begin{equation*}
\#\{\substack{\text{matchings with} \\ \text{$c$ loops}}\}=\sum_{\substack{\text{partitions of $[L]$}\\ \text{into $c$ blocks}}}\ \prod_{b}(r_b-1)!\,2^{r_b-1}
\end{equation*}
Notice that $\prod_b2^{r_b-1}=2^{L-c}$ and that $\sum\prod_b(r_b-1)!$ counts the permutations of $L$ objects with $c$ cycles, i.e.\ the unsigned Stirling number of the first kind $|s(L,c)|$. Hence, $P(c)=2^{L-c}|s(L,c)|/(2L-1)!!$.
Equivalently, $c$ is the sum of independent Bernoulli variables. In fact, one can build the matching step by step: start from a dot, follow its site arc (e.g.\ from $2j-1$ to $2j$), and connect the other end to a uniformly random free dot (e.g.\ from $2j$ to $2i-1$); when a loop closes, restart from a new site. When $k$ sites have been visited, $2L-1-2k$ dots are available, one of which closes the loop, so the loop closes with probability $1/(2L-1-2k)$, independently of the previous steps. Thus $c\overset{d}{=}\sum_{k=1}^{L}\mathrm{Bern}\big(\tfrac{1}{2k-1}\big)$.

\begin{proof}
Since $w\neq0$ on a fraction $2^{1-c}$ of the bitstrings, $P(w\neq0)=\E[2^{1-c}]=2\prod_{k=1}^L\big(1-\tfrac{1}{2(2k-1)}\big)=2\prod_{k=0}^{L-1}\tfrac{4k+1}{4k+2}$. Therefore
\begin{equation}
P(u)=\frac{2^{-(u+1)}P_c(u+1)}{\sum_{c'=1}^{L}2^{-c'}P_c(c')}=\frac{4^{L-u-1}|s(L,u+1)|}{\prod_{k=0}^{L-1}(4k+1)},
\end{equation}
where we used $\prod_{k=0}^{L-1}(4k+2)=2^L(2L-1)!!$. Finally, since $w=2^{c-1}$, $\E[w^k]=2^{1-k}\E[2^{(k-1)c}]=2^{1-k}\prod_{j=1}^L\big(1+\tfrac{2^{k-1}-1}{2j-1}\big)$, averaging over the independent Bernoulli variables. \\
\end{proof}

\paragraph{Arc approximation}
First we prove the exact duality between the single arc process and the full matching process, i.e.\ $M^{(\alpha)}\,V_f=V_{K^{(\alpha)}f}$ with $V_f(\pi)=\sum_p f_p \, n_{p}(\pi)$. The proof is direct: 
\begin{align*}
\begin{split}
&\sum_{\pi} M_{\pi' \pi}^{(\alpha)} V_f(\pi)=\sum_p f_p \sum_{\pi} n_{p}(\pi) M_{\pi \pi'}^{(\alpha)} \\ &= \sum_p f_p \sum_{q} K_{pq}^{(\alpha)} n_{q}(\pi') = \sum_{q} (K^{(\alpha)}f)_{q} n_{q}(\pi') = V_{K^{(\alpha)} f}(\pi')  \, ,      
\end{split}
\end{align*}
where we used the fact that $M^{(\alpha)}$, $K^{(\alpha)}$ are symmetric and that $\sum_{\pi} n_{p}(\pi) M_{\pi \pi'}^{(\alpha)} = \sum_{q} K_{pq}^{(\alpha)} n_{q}(\pi')$ since both quantities equal the probability of $p$ being an arc after one time step starting from $\pi'$. The vectors $V_f$ form a subspace $\mathcal S$, spanned by the $\{n_p\}$, and we denote by $Q$ the orthogonal projector onto $\mathcal S$. Since each layer maps $\mathcal S$ to itself and is symmetric, $[M^{\alpha},Q]=0$.\\
Now we define the action of $\sigma \in S_{2L}$ on matchings $\pi=\{(a_i,b_i)\}_{i=1}^L$ as $\sigma\pi=\{(\sigma(a_i),\sigma(b_i))\}_{i=1}^L$, and on vectors as $[R_\sigma V](\pi)=V(\sigma^{-1}\pi)$. Since $R_\sigma n_p=n_{\sigma p}$, $R_\sigma$ maps $\mathcal S$ to itself. Moreover, it clearly preserves the scalar product $\langle V,W\rangle=\sum_\pi V(\pi)W(\pi)$. Hence, $[Q,R_\sigma]=0$. Consider now the subgroup $\mathbb G\subset S_{2L}$ of permutations mapping sites to sites, generated by permutations of the sites and swaps of the two dots within a site ($\mathbb G\simeq S_2\wr S_L$). Under the action of $\mathbb G$, all site pairs and all pairs connecting different sites are equivalent. So the invariant vectors in $\mathcal S$ are the combinations of the number of site arcs $N_1(\pi)=\sum_j n_{(2j-1,2j)}(\pi)$ and of $L-N_1$. Consequently, if $V$ is invariant under $\mathbb G$, so is $QV$, and $QV=\alpha N_1+\beta(L-N_1)$, where $\alpha,\beta$ are fixed by the orthogonality of $V-QV$ to $N_1$ and to $L-N_1$. This formula can be rearranged as $QV=\E_\infty[V]+\kappa_V\big(N_1-\E_\infty[N_1]\big)$, with $\kappa_V=\big(\E_\infty[VN_1]-\E_\infty[V]\,\E_\infty[N_1]\big)/\big(\E_\infty[N_1^2]-\E_\infty[N_1]^2\big)$, where $\E_\infty[\dots]=\langle 1,\dots\rangle/(2L-1)!!$ denotes the average over uniformly random matchings. For example, for the number of loops $V=c$ one gets $\kappa_c=(2L-3)/(2L-2)$. Therefore, the single-arc approximation becomes 
\begin{equation*}
\E_t[V]=\langle V,P_t\rangle\simeq\langle QV,P_t\rangle=\E_\infty[V]+\kappa_V\big(\E_t[N_1]-\E_\infty[N_1]\big) \, .
\end{equation*}
Notice that, since $N_1\in\mathcal S$, $\E_t[N_1]$ can be computed exactly by evolving the single-arc Markov chain; through the formula above, this single quantity then gives an approximation for any site-invariant observable of the full matching process. For example, the loop counting indicator $\delta(c(\pi),c)$ gives $P_t(c)\simeq P_{\infty}(c)+\kappa_c(\E_t[N_1]-\E_\infty[N_1])$, with $\kappa_c\propto P_{\infty}(c-1)|_{L-1}-P_{\infty}(c)|_L$. The initial diffusive behavior of $\E_t[N_1]$ follows from a simple argument. Each arc end performs a random walk whose gate moves by one site, left or right with probability $\frac12$, at every layer, so after $t$ layers its variance is $\simeq t$ in sites, i.e.\ $\sigma^2\simeq4t$ (in dots). Treating the two ends as independent walkers with Gaussian distribution $G_\sigma$, the probability of the arc being a site arc is $\int G_\sigma^2=1/(2\sqrt\pi\sigma)=1/(4\sqrt{\pi t})$, hence $\E_t[N_1]\simeq L/(4\sqrt{\pi t})$ for $1\ll t\ll L^2$.\\
The same construction applies to generic FGS, by replacing matchings $\pi$ with covariance matrices $\Gamma$. For a pure FGS, $\sum_b\Gamma_{ab}^2=1$ for every row $a$; the squared entries $\Gamma_{ab}^2$ can thus be interpreted as a probability over the partners $b$ of $a$. Arc indicators $n_p(\pi)$ are therefore replaced by $\Gamma_p^2$, and uniform averages over matchings become Haar averages over FGS. The space $\mathcal S=\mathrm{span}\{\Gamma_p^2\}$ is again invariant, with the same single-arc chain $K$. Indeed, $\Gamma_p^2$ is quadratic in the state, so its gate average is the same for Haar-random and Clifford matchgates, since the latter form a 3-design~\cite{Wan_2023,Braccia_2025}; as a Clifford gate merely permutes the $\Gamma_p^2$: $\E_{\rm gate}[\Gamma'^2_p]=\E^{\rm Cl}_{\rm gate}[\Gamma'^2_p]=\sum_qK_{pq}\Gamma_q^2$. The site symmetry is unchanged (with swaps within sites taken in pairs, to preserve parity), and $N_1=\sum_j\Gamma_{2j-1,2j}^2=\sum_j\langle Z_j\rangle^2\in\mathcal S$ has the same average $\E_t[N_1]$ as in the Clifford case, since both evolve with $K$ from the vacuum. The only difference is that $\kappa_V$ is now an overlap over Haar-random FGS, i.e.\ $\E_\infty[\dots]=\int d\mu_{\rm H}(\Gamma)\,\dots$ with $\mu_{\rm H}$ the Haar measure on FGS, which can be estimated by sampling.

\vspace{-2 mm}
\paragraph{Proof of the Theorem} Clifford matchgates are matchgate $3$-design~\cite{Wan_2023,Braccia_2025}, so we can evaluate $Z_t$ for Clifford circuits. -- \begin{proof}
\emph{Step 1.} A state with $c$ loops has $w=2^{c-1}$ on a fraction $2^{1-c}$ of the even strings and $w=0$ elsewhere, hence $\E_x[w^2]=2^{c-1}$. Each site arc is a loop by itself, so $c\ge N_1$, and by Jensen's inequality $Z_t=\E_t[2^{c-1}]\ge2^{\E_t[c]-1}\ge2^{\E_t[N_1]-1}$. \emph{Step 2.} Let $v_t(p)=P_t(p\in\pi)$, so that $\sum_pv_t(p)=L$ and $\E_t[N_1]=\langle\mathbf 1_S,v_t\rangle$, where $S$ is the set of site pairs and $\langle f,g\rangle=\sum_pf(p)g(p)$. The layers $K^{(0)},K^{(1)}$ are symmetric and idempotent, so $A=K^{(0)}K^{(1)}K^{(0)}$ is symmetric, with $A^n=K^{(0)}(K^{(1)}K^{(0)})^n$ and hence $v_{2n+1}=A^n\mathbf 1_S$. For odd $t$ this gives $\E_{2t-1}[N_1]=\langle\mathbf 1_S,A^{t-1}\mathbf 1_S\rangle=\|A^{(t-1)/2}\mathbf 1_S\|^2=\|v_t\|^2$. If a `mass' $m=\sum_{p\in R}v_t(p)$ lies on a set $R$ of pairs, Cauchy--Schwarz gives $\|v_t\|^2\ge m^2/|R|$. We take $R$ to be the pairs whose ends are at most $r=4\sqrt t$ sites apart, so that $|R|\le2L(2r+1)$, and show that $m\ge3L/4$. \emph{Step 3.} Each end of an arc, taken alone, moves as a random walker. Let $\sigma_\tau=\pm1$ according to whether layer $\tau$ leaves the end on the right or the left site of its gate; these coins are fair and independent, also when both ends share the gate. The position of the end then obeys $x_{\tau+1}=x_\tau+\frac12(\sigma_\tau+\sigma_{\tau+1})$, with $\sigma_0=+1$ ($-1$) if the end starts on the left (right) site of its first gate. Summing, $x_t-x_0=\tfrac12\sigma_0+\sum_{\tau=1}^{t-1}\sigma_\tau+\tfrac12\sigma_t$, so $\E[(x_t-x_0)^2]=\frac14+(t-1)+\frac14\le t$; the boundaries only reflect the walk, so the bound still holds. The two ends of the arc starting on site $j$ start on the same site, so its length $\ell^{(j)}_t=x_t-x'_t$ obeys $\E[(\ell^{(j)}_t)^2]\le2\E[(x_t-x_0)^2]+2\E[(x'_t-x_0)^2]\le4t$. By the Chebyshev inequality, $P(|\ell^{(j)}_t|>r)\le\frac14$. Hence $m=\sum_{j=1}^LP(|\ell^{(j)}_t|\le r)\ge3L/4$. \emph{Conclusion.} Steps 2,3 give $\E_{2t-1}[N_1]\ge L/(C'\sqrt t)$, which extends to all depths since one layer keeps a site arc with probability $\ge\frac13$. With Step 2 and $Z_\infty\le\frac12\sqrt{\pi(L+1)}$, $Z_t\le qZ_\infty$ then requires $t\ge L^2/\big(C\log_2^2(qL)\big)$. \\
\end{proof}

\newpage 

\section{Supplemental Material}
\subsection{Large $L$ expansion at fixed $\nu=N/L$ (number preserving random FGS)}

The cumulants of $\ln B$, with $B\sim\Betad(a,b)$, are $\frac{d^n}{dt^n}\ln\E[B^t]\big|_{t=0}=\psi^{(n-1)}(a)-\psi^{(n-1)}(a+b)$.
With $X_j=\ln B_j$, $B_j\sim\Betad(j,\Nh)$, we have $\kappa_n(X_j)=\psi^{(n-1)}(j)-\psi^{(n-1)}(j+\Nh)$. Since $\ln w=\ln D+\sum_{j=1}^N X_j$ with independent $X_j$, and cumulants add over independent variables, we find $\E\ln w=\ln D+\sum_{j=1}^N\big[\psi(j)-\psi(j+\Nh)\big]$ and $\mathrm{Var}\ln w=\sum_{j=1}^N\big[\psi'(j)-\psi'(j+\Nh)\big]$. For $n\ge2$ we use $\psi^{(n-1)}(x)=(-1)^n(n-1)!\sum_{m\ge0}(x+m)^{-n}$. \\

\emph{Mean.} We apply the identity $\sum_{j=1}^M\psi(j)=M\psi(M+1)-M$ to the sums $\sum_{j=1}^N\psi(j)$ and
$\sum_{j=1}^N\psi(j+\Nh)=\sum_{k=1}^{L}\psi(k)-\sum_{k=1}^{\Nh}\psi(k)$. Using also $\ln D=\ln L!-\ln N!-\ln\Nh!$, we obtain
$\E\ln w=f(L)-f(N)-f(\Nh)$, with $f(M)=\ln M!-M\psi(M+1)$. By Stirling, $\ln M!=M\ln M-M+\frac12\ln(2\pi M)+O(1)$, and
$M\psi(M+1)=M\ln M+\frac12+O(1)$, so $f(M)=-M+\frac12\ln(2\pi M)-\frac12+O(1)$. Since $N+\Nh=L$, the terms $-M$ cancel and
$\E\ln w=\frac12-\frac12\ln(2\pi\Lambda)+O(1)$. \\

\emph{Variance.} In the first sum we use $\psi'(x)=\sum_{m\ge0}(x+m)^{-2}$ and group the terms by $k=j+m$: each $k$ occurs
$\min(k,N)$ times, so $\sum_{j=1}^N\psi'(j)=\sum_{k\le N}\frac1k+N\sum_{k>N}\frac1{k^2}=\ln N+\gamma_E+1+O(1)$.
In the second, all arguments exceed $\Nh\propto L$, so $\psi'(x)=\frac1x+O(x^{-2})$ gives
$\sum_{j=1}^N\psi'(j+\Nh)=\sum_{k=\Nh+1}^{L}\frac1k+O(1)=\ln\frac{L}{\Nh}+O(1)$.
Subtracting, $\mathrm{Var}\ln w=\ln\Lambda+1+\gamma_E+O(1)$. \\

\emph{Gaussian limit.} Since the $X_j$ are independent but not identically distributed, we use the central limit theorem in
Lyapunov's form: the standardized sum converges to $\mathcal N(0,1)$ provided the fluctuations of single terms are negligible compared with those of the sum, or more specifically
$\sum_{j=1}^N \E\big(X_j-\E X_j\big)^4 / \Big(\sum_{j=1}^N\mathrm{Var}\,X_j\Big)^2 \longrightarrow 0$.
From the truncated series above, $|\kappa_n(X_j)|\le(n-1)!\sum_{m\ge0}(j+m)^{-n}$, a bound independent of
$\Nh$ and $L$. Bounding the terms with $m\ge1$ by $\int_j^\infty x^{-n}dx=j^{1-n}/(n-1)$ gives
$|\kappa_n(X_j)|\le n\,(n-2)!\,j^{1-n}$; in particular $\mathrm{Var}\,X_j\le 2/j$ and $|\kappa_4(X_j)|\le8/j^3$.
Hence $\E(X_j-\E X_j)^4=\kappa_4(X_j)+3\kappa_2(X_j)^2=O(j^{-2})$ has a bounded sum over $j$, while the denominator
$(\mathrm{Var}\ln w)^2\simeq(\ln\Lambda)^2$ diverges. Lyapunov's condition holds, and $z=(\ln w-\E\ln w)/\sqrt{\mathrm{Var}\ln w}\to\mathcal N(0,1)$. The convergence is slow: the third cumulant of $\ln w$ tends to $-2\zeta(2)=-\pi^2/3$, so the skewness of $z$ is $-(\pi^2/3)(\ln\Lambda)^{-3/2}$ to leading order. \\

\subsection{Large $L$ expansion (generic random FGS)}

As before, $\ln w=\ln D+\sum_{j=1}^{L-1}X_j$ with independent $X_j=\ln B_j$, $B_j\sim\Betad(j,j)$, and $D=2^{L-1}$, so that
$\kappa_n(\ln w)=\delta_{n,1}\ln D+\sum_{j=1}^{L-1}\kappa_n(X_j)$ with $\kappa_n(X_j)=\psi^{(n-1)}(j)-\psi^{(n-1)}(2j)$. For $n\ge2$ the series of $\psi^{(n-1)}$ truncates,
$\kappa_n(X_j)=(-1)^n(n-1)!\sum_{m=0}^{j-1}(j+m)^{-n}$. \\

\emph{Mean.} The duplication formula $\psi(2j)=\frac12\psi(j)+\frac12\psi(j+\frac12)+\ln2$ gives
$\psi(j)-\psi(2j)=\frac12\big[\psi(j)-\psi(j+\frac12)\big]-\ln2$, and the terms $-\ln2$ cancel $\ln D$ exactly:
$\E\ln w=\frac12\sum_{j=1}^{L-1}\big[\psi(j)-\psi(j+\frac12)\big]$. From $\psi(x+1)=\psi(x)+\frac1x$ one gets
$\sum_{j=1}^{M}\psi(j+a)=(M+a)\psi(M+a+1)-a\,\psi(a+1)-M$. Applying it with $M=L-1$ and $a=0,\frac12$,
$\E\ln w=\frac12\big[M\psi(M+1)-(M+\frac12)\psi(M+\frac32)+\frac12\psi(\frac32)\big]$.
For large $M$, $M\psi(M+1)=M\ln M+\frac12+o(1)$ and $(M+\frac12)\psi(M+\frac32)=M\ln M+\frac12\ln M+1+o(1)$. With
$\psi(\frac32)=2-\gamma_E-2\ln2$, this gives $\E\ln w=-\frac14\ln(4L)+\frac{1-\gamma_E}4+o(1)$. \\

\emph{Variance.} We group the terms of $\mathrm{Var}\ln w=\sum_{j=1}^{L-1}\sum_{m=0}^{j-1}(j+m)^{-2}$ by $k=j+m$, for which
$k/2<j\le\min(k,L-1)$. For $k\le L-1$, $k$ occurs $\lceil k/2\rceil=\frac k2+\frac12[k\text{ odd}]$ times, giving
$\sum_{k\le L-1}\frac1{2k}+\frac12\sum_{k\text{ odd}}\frac1{k^2}=\frac12(\ln L+\gamma_E)+\frac{\pi^2}{16}+o(1)$.
For $L-1<k<2L-2$, $k$ occurs $\simeq L-\frac k2$ times, giving the Riemann sum
$\sum_k\frac{L-k/2}{k^2}\to\int_1^2\frac{1-x/2}{x^2}\,dx=\frac12-\frac12\ln2$.
Adding, $\mathrm{Var}\ln w=\frac12\ln\frac L2+\frac{1+\gamma_E}2+\frac{\pi^2}{16}+o(1)$. \\

\emph{Gaussian limit.} As in the number-preserving case, we use Lyapunov's condition. The truncated series has $j$ terms, each at most $j^{-n}$, so $|\kappa_n(X_j)|\le(n-1)!\,j^{1-n}$. Hence $\E(X_j-\E X_j)^4=O(j^{-2})$ has a bounded sum over $j$, while $(\mathrm{Var}\ln w)^2$ diverges, and $z=(\ln w-\E\ln w)/\sqrt{\mathrm{Var}\ln w}\to\mathcal N(0,1)$. The third cumulant of $\ln w$ stays bounded, so the skewness of $z$ decays slowly, as $(\ln L)^{-3/2}$.

\end{document}